\documentclass{pasj00}
\usepackage[]{natbib,graphicx}

\newcommand{\hinode}{{\em Hinode{}}}
\newcommand{\trace}{{\em TRACE{}}}

\newcommand{\alfven}{Alfv\'{e}n}
\newcommand{\ca}{Ca~II H}

\begin{document}
\SetRunningHead{De Pontieu et al.}{Two Chromospheric Spicule Types}
\Received{2007/06/10}
\Accepted{accepted for Hinode special issue}
\title{A Tale Of Two Spicules: The Impact of Spicules on the Magnetic Chromosphere}

\author{Bart \textsc{De Pontieu}\altaffilmark{1},
        Scott {\sc McIntosh}\altaffilmark{2,3},        
Viggo H. \textsc{Hansteen}\altaffilmark{4,1},
        Mats {\sc Carlsson}\altaffilmark{4}}
\email{ bdp@lmsal.com, mscott@hao.ucar.edu, Viggo.Hansteen@astro.uio.no,Mats.Carlsson@astro.uio.no}
\author{        C.J. {\sc Schrijver}\altaffilmark{1},
                T.D. {\sc Tarbell}\altaffilmark{1},
        A.M. {\sc Title}\altaffilmark{1},
        R.A. {\sc Shine}\altaffilmark{1}}
\email{schryver@lmsal.com,  tarbell@lmsal.com, title@lmsal.com, shine@lmsal.com}
\author{Y. {\sc Suematsu}\altaffilmark{5},
        S. {\sc Tsuneta}\altaffilmark{5},
        Y. {\sc Katsukawa}\altaffilmark{5},
        K. {\sc Ichimoto}\altaffilmark{5},
        T. {\sc Shimizu}\altaffilmark{6}
        S. {\sc Nagata}\altaffilmark{7}}
\email{yoshinori.suematsu@nao.ac.jp,saku.tsuneta@nao.ac.jp, yukio.katsukawa@nao.ac.jp, kiyoshi.ichimoto@nao.ac.jp, 
toshifumi.shimizu@nao.ac.jp, nagata@kwasan.kyoto-u.ac.jp}

\altaffiltext{1}{Lockheed Martin Solar and Astrophysics Laboratory, Palo Alto, CA 94304, USA}
\altaffiltext{2}{High Altitude Observatory, National Center for Atmospheric Research, PO Box 3000, Boulder, CO
80307, USA}
\altaffiltext{3}{Department of Space Studies, Southwest Research Insititute, 1050 Walnut St, Suite 400, Boulder, CO
80302, USA}
\altaffiltext{4}{Institute of Theoretical Astrophysics, University of Oslo, PB 1029 Blindern, 0315 Oslo Norway}
\altaffiltext{5}{National Astronomical Observatory of Japan,  Mitaka,
  Tokyo, 181-8588, Japan}
\altaffiltext{6}{ISAS/JAXA, Sagamihara, Kanagawa, 229-8510, Japan}
\altaffiltext{7}{Kwasan and Hida Observatories, Kyoto University,Yamashina, Kyoto, 607-8471, Japan}

\KeyWords{Sun: Chromosphere \-- Sun: Transition Region} 

\maketitle

\begin{abstract}
   We use high-resolution observations of the Sun in \ca{}
  (3968\AA) from the Solar Optical Telescope on \hinode{} to show that
  there are at least two types of spicules that dominate the structure
  of the magnetic solar chromosphere. Both types are tied to the
  relentless magnetoconvective driving in the photosphere, but have
  very different dynamic properties. ``Type-I'' spicules are driven by
  shock waves that form when global oscillations and convective flows
  leak into the upper atmosphere along magnetic field lines on 3-7
  minute timescales. ``Type-II'' spicules are much more dynamic: they
  form rapidly (in $\sim$10s), are very thin ($\le$200km wide), have
  lifetimes of 10-150s (at any one height) and seem to be rapidly
  heated to (at least) transition region temperatures, sending
  material through the chromosphere at speeds of order 50-150 km/s.
  The properties of Type II spicules suggest a formation process that
  is a consequence of magnetic reconnection, typically in the vicinity
  of magnetic flux concentrations in plage and network. Both types of
  spicules are observed to carry Alfv\'en waves with significant
  amplitudes of order 20 km/s.
\end{abstract}

\section{Introduction}
The dynamics of the magnetized chromosphere are dominated by spicules
\citep[at the limb, e.g.,][]{Beckers1968} and related flows such as
mottles and fibrils \citep[on the disk, see
e.g.,][]{Hansteen+etal2006,DePontieu+etal2007a,DePontieu+etal2007b}.
Spicular features are also visible at the limb in many spectral lines
formed at transition region (TR) temperatures \citep{Mar92,Wil00}, and
recent observations suggest that much of the TR and perhaps coronal
dynamics and energetics are intimately linked to spicule-like jets
\citep{McIntosh+etal2007}.

Spicules have been studied for many decades, but until recently they
have remained difficult to understand because their structure and
dynamics were too close to observational resolution limits
\citep{Sterling2000}. Recently, high spatial and temporal resolution
observations from the ground \citep[Swedish 1 m Solar Telescope --
SST,][]{scharmer2003SST} and advanced radiative MHD simulations
\citep[e.g.,][]{Hansteen2005,Hansteen+etal2006,Hansteen+Carlsson+Gudiksen2007}
have allowed us to make significant progress in our understanding of
mottles and fibrils
\citep{Hansteen+etal2006,DePontieu+etal2007a,Rouppe+etal2007,Heggland+etal2007}.
Comparisons between the observations and simulations showed that
active region dynamic fibrils and a subset of quiet Sun mottles are
formed when photospheric oscillations and convective flows leak into
the chromosphere along magnetic field lines, where they form shocks to
drive jets of chromospheric plasma, as suggested by
\citet{DePontieu+etal2004}.

While our understanding of fibrils and mottles (both observed on the
disk) has improved considerably, it remains unclear how or whether
these features are related to spicules at the limb
\citep[e.g.,][]{Tsiropoula+etal1994,Suematsu+etal1995}.  Here we use
high-resolution magnetograms, and optical and EUV images and spectra
from {\it Hinode} \citep[][]{Ichimoto2005a,Ichimoto2005b}, as well as
1600\AA{} passband imaging from the Transition Region and Coronal
Explorer \citep[{\em TRACE};][]{Handy1999}, to demonstrate that there
are (at least) two types of chromospheric spicules. Both types are
intrinsically tied to magneto-convective field evolution but are
readily distinguished by their very different dynamic behavior.

Section~\ref{analysis} will detail the observations and analysis
techniques used in this study. Sect.~\ref{discuss} discusses the
results of the analysis and the potential impact of our results on our
understanding of the transition region (and corona). We also pose
several unresolved issues and questions in this section.

\section{Observations \& Data Analysis}\label{analysis}

We use a series of image sequences obtained since November 2006 with
the SOT Broadband Filter Imager \citep[BFI, ][]{Tarbell+etal2007} in the
\ca{} 3968\AA{} wavelength range. The \ca{} filter is fairly broad
with a FWHM of 2.2 \AA\ so that both photospheric and chromospheric
emissions contribute. \citet{Carlsson+etal2007} show that the
contribution function on the disk is typically dominated by strong
photospheric components, with the middle to upper chromosphere
contributing only a small fraction of the emission. Off the limb, much
of the emission comes from middle to upper chromospheric plasma. The
timeseries we study have been obtained for a variety of locations:
quiet Sun (QS), active region (AR) and coronal hole (CH). They have a
fixed cadence between 4 and 11.2 seconds (with an exposure time of 0.5
s) and a spatial resolution determined by the diffraction limit of the
SOT, of order 0.16\arcsec{} for the \ca{} images. The pixel size is
0.054 \arcsec, with a field of view varying between 1024x2048 and
2048x1024 pixels.  To correct the images for dark current and
miscalleneous camera artefacts (e.g., the two columns in the center
which are missing in the original data), we apply the IDL routine
{\tt fg\_prep.pro}, which is part of the Hinode tree of solarsoft
\citep{Tarbell+etal2007}. While the SOT correlation tracker removes
most of the jitter introduced by the spacecraft, we remove the
remaining slow drift of the time series by performing a rigid
co-alignment of the timeseries. To this end we use cross-correlation
techniques (from image to image) and apply the cumulative offsets to
the whole timeseries (usually about 1 hour long). Using this technique
we achieve datasets that are co-aligned to much better than one pixel
from image to image.

\begin{figure}
  \begin{center}
  \FigureFile(90mm,90mm){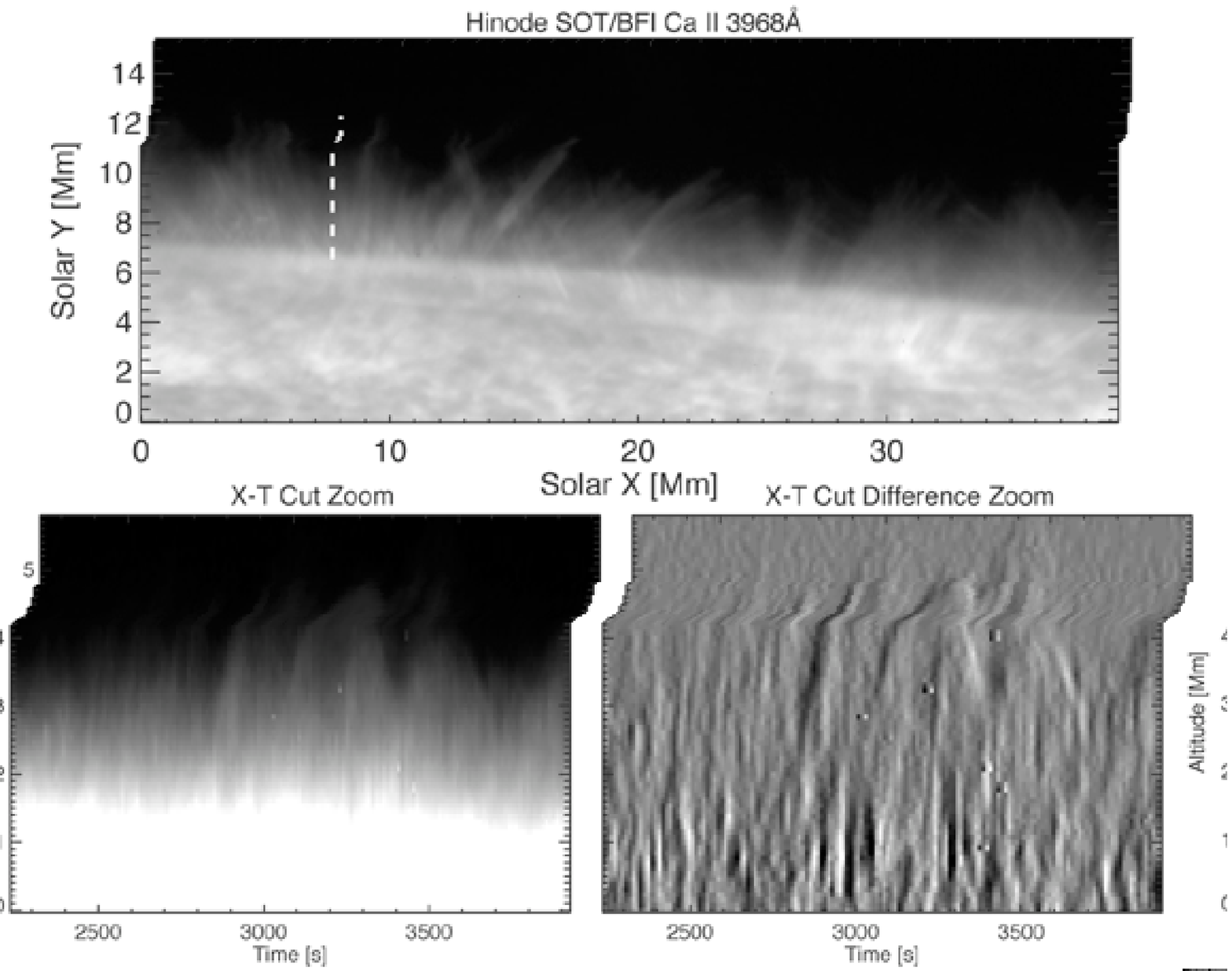}
      \end{center}
\caption{The top panel shows spicules at the quiet Sun limb on November
  22, 2006 in SOT/BFI \ca{} 3968\AA\ data. The bottom panels show
  space-time (``xt'') plots along the location indicated by a dashed
  line in the top panels, for both the original data and time
  differenced data. The space-time plot is dominated by short-lived
  vertical stripes (type II spicules) and longer-lived parabolic paths
  (type I spicules). }
\label{fig1}
\end{figure}

Movies of the limb in \ca{} reveal a wealth of information on the
dynamic solar chromosphere: surges, prominences, macrospicules and
spicules (see Fig. \ref{fig1}). We focus here on the spicules. Our
timeseries show that the limb is dominated by these thin, long and
highly dynamic features.  They have diameters from about 700 km (1
\arcsec) down to the resolution limit of the telescope (120 km). Their
maximum lengths vary from a few hundred kilometers to 10,000 km with
most below 5,000 km.  Many are seen to go up- and down, whereas others
show only upward motion, followed by a rapid fading from the \ca{}
passband. In addition, the majority of spicules seen in our timeseries
undergo significant lateral motion, i.e., motion that appears to be
transverse to the long axis of the spicules. These lateral motions
have been interpreted as Alfv\'enic motions \citep[for more details,
see][]{DePontieu+etal2007c}.

A visual study of the dynamic behavior of the spicules reveals that
there are {\em at least} two different populations governed by very
different timescales. To study these populations in detail, we create
a Fourier filtered timeseries for each individual spatial pixel. The
Fourier filtering is done only in time, and we apply two different
filters.  The first, low-frequency, filter is a Gaussian centered on
3mHz with a width of 0.5 mHz. This filter is designed to isolate
significant propagating 5 minute power off the limb. These low
frequency filtered movies show the ``type I'' spicules: relatively
slowly evolving features that typically move up and down during their
lifetimes of order 3-7 minutes. A second, much more dynamic spicular
component is invisible in these movies because these ``type
II'' spicules often appear or disappear within 4.8 s (which is the
highest cadence we have obtained). The second filter is designed to
bring out these short-lived features. It is a combination of a
high-pass filter ($\ge$15mHz) with a low-pass filter placed at a
frequency which is 2mHz lower than the Nyquist frequency of the
timeseries studied. The low-pass filter is designed to remove any
effects on the filtered timeseries from any residual jitter not
removed by our co-alignment procedure. Movies of high frequency
filtered data isolate the ``type II'' spicules very well, both at the
limb and on the disk, where they occur as ``straws'' predominantly in
and around the chromospheric network and plage. We refer to them as
``straws'' on the disk since we believe the disk features have a close
resemblance to the straws described by \citet{Rutten2006,Rutten2007}.

\begin{figure}
  \begin{center}
     \FigureFile(90mm,90mm){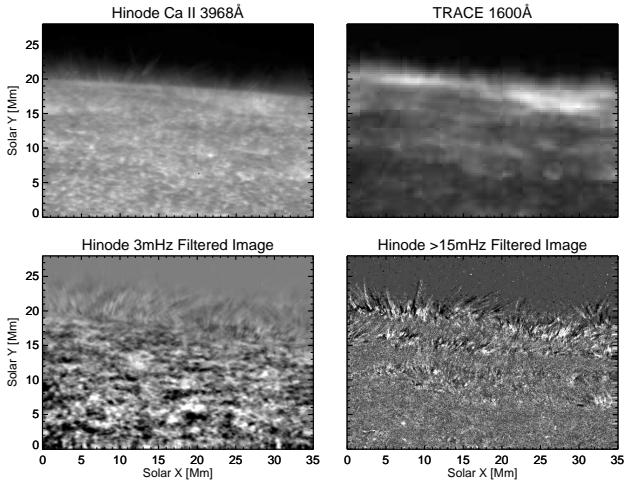}
      \end{center}
\caption{Sample images from the Fourier filtered timeseries of SOT-BFI
  \ca{}~3968\AA{} images (top left) in the 3 (bottom left) and
  $\ge$15mHz (bottom right) passbands. These passbands isolate the two
  spicular species (see text). A comparison of the high frequency
  filtered \hinode{} images with those from the \trace{} 1600\AA{}
  passband suggests that type II spicules and ``straws'' (on the disk)
  are associated with bright C~IV transition region emission (top right).}
\label{fig2}
\end{figure}

Figure \ref{fig2} shows the comparison of a \ca{} 3968\AA{} image (top
left) with the corresponding frame from the low (bottom left) and high
frequency (bottom right) filtered timeseries.  For later reference we
compare these images with a simultaneous image from the 1600\AA{}
passband of Transition Region and Coronal Explorer
\citep[TRACE;][]{Handy1999}. 

The following subsections discuss the appearance of each spicule type
and provide some of their basic properties.

\subsection{Type-I Spicules}

Type I spicules are dominated by dynamics on timescales of 3-7 min.
They show a succession of upward and downward motion. While many
undergo transverse motions during their lifetime, xt-cuts with x
roughly perpendicular to the limb show that type I spicules that do
not move transversely often undergo parabolic paths (bottom left panel
of Fig. \ref{fig1} and top panel of Fig. \ref{fig4}), most often with
a deceleration that is not equal to solar gravity (i.e., their paths
are non-ballistic, with decelerations between 50 and 400 m s$^{-2}$).
This behavior is identical to that of AR dynamic fibrils and some QS
mottles \citep{Suematsu+etal1995} as observed in H$\alpha$ with the
Swedish 1m Solar Telescope \citep{Hansteen+etal2006,
  DePontieu+etal2007a, Rouppe+etal2007}.  These authors measured the
parabolic parameters of these fibrils/mottles and found a linear
correlation between the deceleration and maximum velocity of the paths
taken by the fibrils and mottles. This linear correlation can be
readily understood in terms of shock wave physics
\citep{DePontieu+etal2007b,Heggland+etal2007}.

Using the same technique as described in detail in
\citet{DePontieu+etal2007a}, we measured the (constant) deceleration
and maximum velocity of 20 spicules that followed a parabolic path, as
observed at the limb in SOT/BFI \ca{} timeseries. We find that the
observed decelerations and maximum velocities of the type I spicules
(red triangles in Fig. \ref{fig3}) cover a similar range as for the
fibrils (small black asterisks), quiet Sun mottles (blue diamonds) and
jets in the numerical simulations (bottom panel Fig. \ref{fig3}). In
addition, they show an identical linear correlation between
deceleration and maximum velocity, as illustrated in Figure
\ref{fig3}. Given these similarities and the linear correlation, it
thus seems highly likely that the shockwave-driven mechanism also
powers this subset (type I) of spicules observed in \ca{} at the limb.
In other words, type I spicules seem to be the limb equivalent of
dynamic fibrils in active regions, and a subset of quiet Sun mottles.
 
In fact, on the disk these type I spicules are sometimes visible in
absorption in Hinode/SOT \ca{} data as well, but only if the
background is bright enough (plage or bright network, especially
towards the limb). The broad SOT \ca{} filter with its large
photospheric contributions decreases the visibility on the disk of
these dark, absorbing features sometimes, but similar behavior has
been seen more clearly in narrow-band \ca{} core images taken with the
SST.

Type I spicules seem to be quite dominant at the active region limb,
but their visibility in quiet Sun and coronal holes varies. Especially
in coronal holes they are more difficult to detect in SOT/BFI \ca{},
although sometimes they seem to appear there as short and dark ($\sim
2$Mm) absorbing features against a background of bright (and longer)
type II spicules.


\subsection{Type-II Spicules}

Space-time plots (``xt-cuts'') of Hinode/SOT \ca{} timeseries at the
limb (bottom panel of Fig. \ref{fig4}) and close to the network or
plage reveal a second type of spicules. These type II spicules are
highly dynamic, develop apparent speeds between 50-150 km/s, reach
lengths between 1,000 and 7,000 km and often disappear over their
whole length within one or a few timesteps (5-20 s). Their lifetimes
at any one height are usually between 10 and 60 s as illustrated in
the bottom panel of Fig. \ref{fig5}. Their duration at any one height
shows a Gaussian distribution centered at 45 s with a width of 20 s in
coronal holes. In quiet Sun they are somewhat shorter-lived, with an
average around 35 s at any one height. The apparent upward speeds of
type II spicules range between 40 km/s and 300 km/s, with the bulk
between 50-150 km/s (top panel of Fig. \ref{fig5}). This is
significantly higher than type I spicules, which typically do not
reach maximum velocities higher than 40 km/s. Space-time plots of type
II spicules reveal that a significant number appear to be slower in
during very short initial phase (see bottom panel of Fig. \ref{fig4})
and seem to accelerate as they reach greater heights towards the end
of their short life.

Individual type II spicules are often weak and show up as brightness
enhancements (of order 5-10\% compared to the background) above a
background intensity that typically shows an exponential drop-off with
height with scale heights between 2 Mm (AR and QS) and 3 Mm (coronal
hole). Type II spicules appear to travel upwards because of their
rapid disappearance or fading at the end of their lifetime.  This is
clearly visible in Fourier filtered movies using the high frequency
filter described before.  Given the rapid appearance and
disappearance, these type II spicules also dominate difference movies
made by subtracting neighboring images from timeseries with the
highest possible cadence.

Unsharp masking of SOT/BFI \ca{} images at scales of 150-200 km
reveals that most type II spicules are very thin: the line-of-sight is
dominated by a myriad of these features. It may well be that observed
background intensity is a superposition of many type II spicules.
Like type I spicules, most type II spicules undergo significant
transverse motions \citep{DePontieu+etal2007c}.

Type II spicules are typically shorter in ARs, especially when seen on
the disk as straws: they often do not become longer than 1-2 Mm. In QS
they routinely reach lengths of order several Mm, and they are tallest
in coronal holes. Type II spicules dominate coronal holes: many reach
heights of 5,000 km or more. As mentioned before, type I spicules seem
to be minimally present in CHs.

What is the ultimate fate of these type II spicules and/or straws?
Their extremely rapid disappearance from the Hinode/SOT \ca{} passband
suggests that ionization of Ca+ ions caused by strong heating could be
occurring. This is tentatively illustrated by a preliminary comparison
between the location of type II spicules and straws in Hinode/SOT
\ca{} (using the high frequency filtered data) with C IV emission
(formed at 100,000 K) in TRACE 1600\AA. Figure \ref{fig2} shows that
the highest density of straws and type II spicules (bottom right
panel) occurs where TRACE 1600 \AA\ shows the largest excess
brightness compared to the unfiltered SOT/BFI \ca{} image (upper right
panel). Since both the TRACE 1600 \AA\ passband and the SOT \ca{}
passband are mostly dominated by photospheric or low chromospheric
contributions, the excess brightness of TRACE 1600 \AA\ points towards
a significant contribution of C IV emission at the location where the
straws are the strongest and most dense. This suggests that straws and
type II spicules may be heated to TR temperatures of at least 100,000
K.  Comparisons between Hinode/SOT \ca{} images and TRACE or
Hinode/EIS-XRT spectra and images will be necessary to shed further
light on the ultimate fate of type II spicules.

\section{Discussion}\label{discuss}

The presence of at least two different spicule types indicates that
there are several different mechanisms working on the Sun to produce
spicules. This complicated picture has contributed significantly to
the multitude of theoretical models and general confusion in the past
regarding the cause of spicules \citep[see, e.g.][]{Sterling2000}.
For example, there have been many conflicting reports in the
literature on whether spicules fade from view or not
\citep{Beckers1968}. This discussion becomes much clearer when taking
into account the different properties of both spicule types: type I's
move up and down, whereas type II's fade.

The first spicule driving mechanism is now fairly well understood.
Photospheric oscillations and convective motions can leak into the
chromosphere along magnetic flux concentrations (i.e., low plasma
$\beta$ environment), where they form shock waves that drive jets of
plasma upwards \citep{DePontieu+etal2004}. These jets fall back down
after a few minutes, so that the top of the jet tracks a parabolic
path. The constant deceleration and maximum velocity of the parabolic
motion are linearly correlated because of shockwave physics
\citep{Heggland+etal2007}. This mechanism seems to drive the type I
spicules we see here and also causes much of the dynamics seen in
H$\alpha$ on the disk around and above plage regions
\citep{Hansteen+etal2006,DePontieu+etal2007a}. It has also been
associated with a subset of quiet Sun mottles seen in H$\alpha$ on the
disk \citep{Rouppe+etal2007}.


Advanced radiative MHD simulations in two and three dimensions
reproduce the observed parabolic tracks, decelerations and maximum
velocities very well \citep{Hansteen+etal2006,DePontieu+etal2007a}.
They show that velocities of order 10-40 km/s can be expected, and
that these jets do not get heated out of the \ca{} passband
\citep{Langangen+etal2007}. The lifetime of these jets is set by the
inclination of the magnetic field in the photosphere and low
chromosphere: significantly inclined field lowers the acoustic
cutoff frequency and allows leakage of the underlying wave spectrum of
the photosphere \citep[the dominant $\sim 5$ min of the p-modes, see
also][]{Jefferies+etal2006,McIntosh+Jefferies2006}. Regions where the
field is more vertical will be dominated by the chromospheric acoustic
cutoff frequency (3 min): the energetic p-modes cannot leak upward,
with shorter and shorter-lived type I spicules as a result. The effect
of the inclination of the magnetic field on the wave leakage may help
explain why type I spicules are difficult to observe in coronal holes.
The magnetic field in coronal holes is more unipolar so that the
overall direction of the field is more vertical (since it lacks much
of the opposite polarity in close proximity that both QS and AR have).
This could lead to less leakage of p-modes, and shorter, less
energetic type I spicules as a result. It is also possible that type I
spicules are harder to observe because of radiative transfer effects,
caused by the different thermodynamic conditions in coronal holes.

\begin{figure}
  \begin{center}
    \FigureFile(80mm,55mm){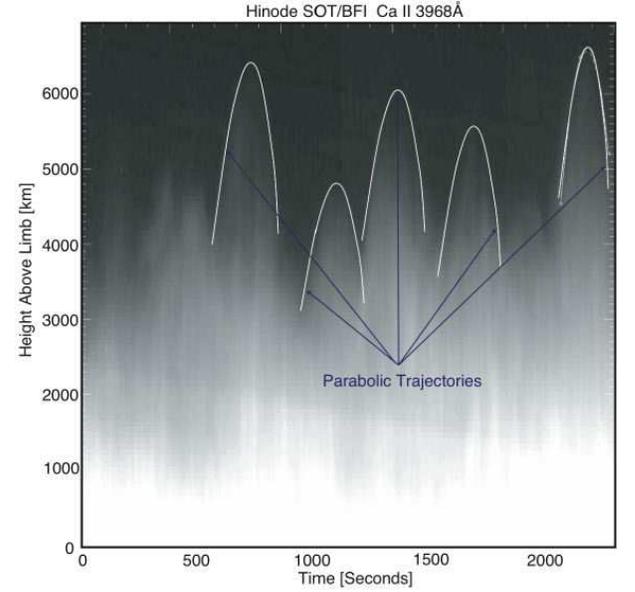}    
    \FigureFile(87mm,58mm){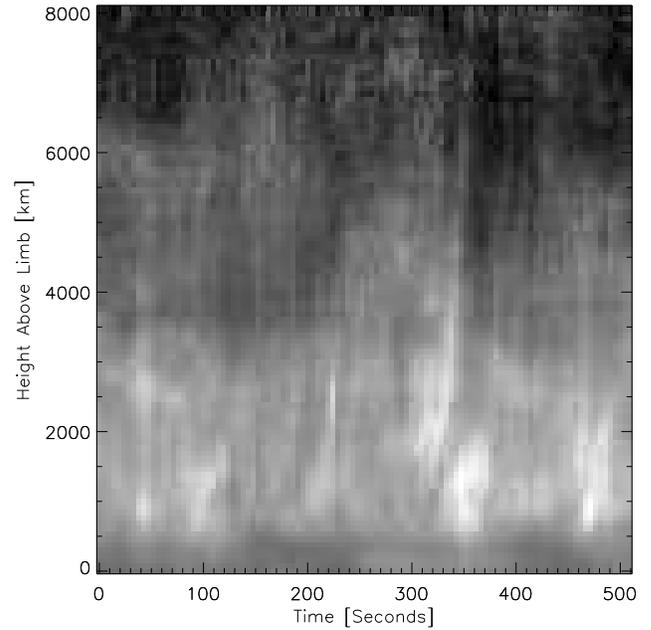}
    %
   \end{center}
\caption{The top panel shows a succession of parabolic paths taken by
  type I spicules in a space-time plot of Hinode/SOT \ca{} 3968\AA\ 
  data with the spatial direction perpendicular to the quiet Sun limb.
  The bottom panel shows a space-time plot of short-lived and fast
  type II spicules at the coronal hole limb. At each height, the data
  in the bottom panel has been divided by the average brightness for
  that height to bring out more details at the top of the spicules.
  Typical for type II spicules are short-lived vertical tracks, some
  of which shown signs of acceleration (e.g., $t=320$ s). This figure
  is accompanied by two movies that illustrate the different dynamics
  of both spicule types. Movie 1 shows the temporal evolution of \ca{}
  3968\AA\ images of the limb dominated by quiet Sun and weak active
  region with a cadence of 8 s, as taken with Hinode/SOT-BFI on 22
  November 2006. Many of the spicules shown in movie 1 are type I
  spicules, with up- and down motion along parabolic paths. Movie 2
  shows \ca{} 3968\AA\ images of a coronal hole on 19 March 2007. Most
  spicules seen here are type II.}
\label{fig4}
\end{figure}

\begin{figure}
  \begin{center}
    \FigureFile(80mm,140mm){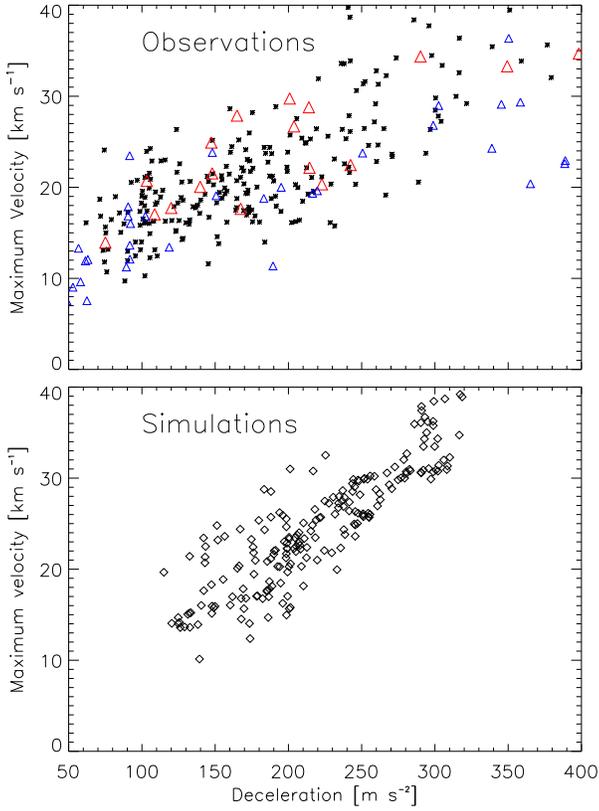}
   \end{center}
\caption{Scatter plot of maximum velocity vs. deceleration for the
  parabolic paths of type I spicules (from Hinode/SOT \ca{} data, red
  triangles), quiet Sun mottles (from SST data, blue triangles) and
active region fibrils (small black asterisks). The fibril data was
corrected for line-of-sight projection by taking into account the
viewing angle based from a magnetic field extrapolation \citep[for
details, see ][]{DePontieu+etal2007a}. Type I spicules shows the same
linear correlation as fibrils and quiet Sun mottles \citep[the latter
data is taken from, ][]{Rouppe+etal2007}. The bottom panel shows a
similar correlation for parabolic parameters of jets from numerical
simulations \citep{Hansteen+etal2006}. These jets are caused by shock
waves that form when oscillations and convective motions leak into the
chromosphere. The scatter in the observational data is caused by the
uncertainty in line-of-sight projection which is unknown for the
mottles and type I spicules, and only roughly known for the fibrils.}
\label{fig3}
\end{figure}

\begin{figure}
  \begin{center}
    \FigureFile(80mm,140mm){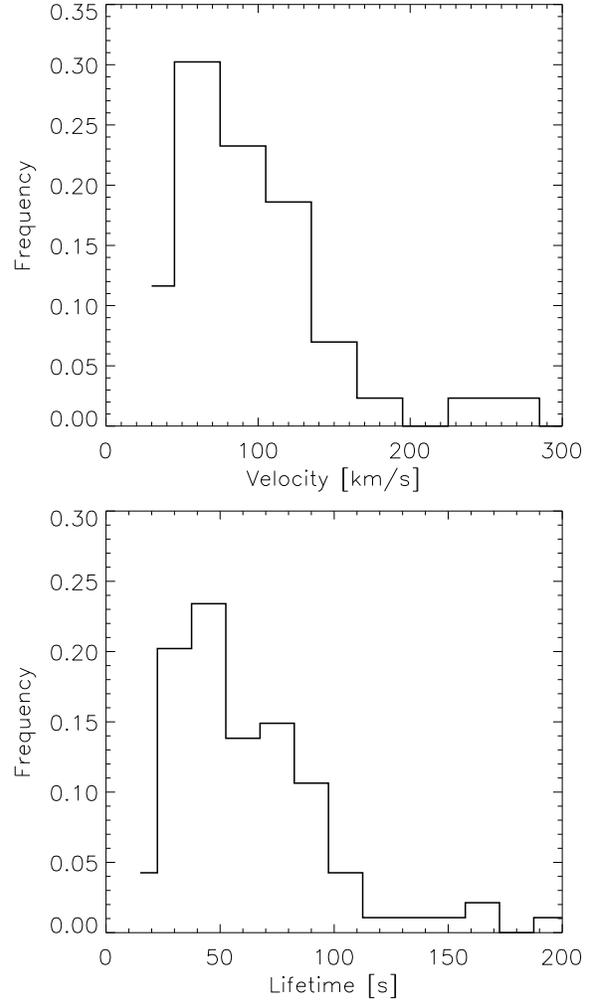}
   \end{center}
\caption{Histogram of upward velocities and lifetimes of type II spicules in
  a coronal hole, from \ca{} 3968\AA\ data taken with SOT/BFI on 19
  March 2007. The average lifetime is of order 45 s, with upward
  velocities along the type II spicules ranging between 40 and 200
  km/s.}
\label{fig5}
\end{figure}

Comparisons with numerical simulations show that the mechanism that
drives type I spicules does not produce jets with (real or apparent)
speeds above 50 km/s or rapid fading caused by heating to TR
temperatures such as we seem to observe in the type II spicules
\citep{Hansteen+etal2006,DePontieu+etal2007a,Langangen+etal2007}.  So
what drives type II spicules and/or straws? An obvious candidate for
the driver of type II spicules is reconnection, which has been invoked
as a driver for spicules in the past, usually as a result of the
interaction between photospheric flux concentrations of opposite
polarity \citep[e.g.,][]{Uch69,Tak00,Moo99,
  Tsiropoula2004,Tziotziou+etal2003,Tziotziou+etal2004,Sterling2000}.
Perhaps reconnection of network and/or plage fields with the turbulent
magnetic field on granular scales \citep[for which there is now
significant evidence, e.g., ][]{Trujillo-Bueno+etal2004,Lites2007} is
driving much of the type II spicular activity? Some of the taller jets
we see in our \ca{} data seem to be associated with twisting motions,
which may be indicative of this type of reconnection. However, it is
not clear at this stage how dominant these kinds of events are at the
limb.  In addition, it is not (yet?) clear whether such a scenario
could explain the presence of straws in the center of strong plage
regions, which are observed to be mostly unipolar with Hinode/SOT-NFI
magnetograms. Perhaps these straws (and some type II spicules) are
caused by the dissipation of currents (and subsequent heating of
plasma) caused by the continual braiding of the field? It is
interesting to note that our space-time plots of type II spicules
(bottom panel, Fig. \ref{fig4}) show strong similarities with similar
cuts along so-called penumbral ``micro-jets'' described by Katsukawa
et al. in this volume. The latter seem to be caused by reconnection at
tangential discontinuities of the magnetic field between neighboring
penumbral filaments with similar polarity, but different field
orientations. Such discontinuities can also be expected above plage
regions because the relentless magnetoconvective forcing of
photospheric flux concentrations shuffles the magnetic field
significantly, especially at greater heights \citep[see,
e.g.][]{vanBallegooijen+etal1998}. It is not yet fully clear whether
this formation mechanism is compatible with the geometry and apparent
field-aligned flows of straws/type II spicules \citep[see,
e.g.][]{Tarbell+etal1999}, especially given the apparent heating of
plasma along the whole spicule. Detailed numerical studies
\citep[along the lines of, e.g., ][]{Sterling+etal1993} with more
advanced chromospheric radiative transfer calculations will be
necessary to determine what kind of heat deposition (and over which
height range) is necessary to produce the thermal and dynamic
properties of straws and/or type II spicules. The work of
\citet{Sterling+etal1993} suggests that only a small subset of
solutions in which significant energy is deposited in the middle
chromosphere come close to producing features with speeds of 50-100
km/s that reach TR temperatures above 20,000 K.

The presence of straws has been described before by \citet{Rutten2007}
who used observations on the disk with the Dutch Open Telescope in La
Palma to describe their morphology and association with the network.
How do they relate to features in other chromospheric passbands? We
have also seen straws in H$\alpha$ linecenter observations made with
the SST, although at significantly reduced contrast. In the SST disk
data, the straws are often obscured by fibrils and mottles, which
appear as dark features in H$\alpha$. Preliminary comparisons of SOT
\ca{} and H$\alpha$ linecenter images at the limb show that a majority
of \ca{} features also show up in H$\alpha$, with the exception of
some of the shorter type II spicules that appear at lower heights
where H$\alpha$ linecenter shows little contrast. Detailed analysis of
straws in different chromospheric lines will be necessary to
disentangle the complicated appearance of spicules and their disk
counterparts, which has been a long-standing source of confusion
\citep{Grossmann+Schmidt1992}. Such an analysis will also be able to
pin down whether all of the observed velocities are caused by real
mass motions. This is clearly the case for type I spicules and
fibrils/mottles as shown by comparisons to spectra
\citep{Langangen+etal2007}. Similar work will be necessary for the
extremely thin type II spicules and straws with their high apparent
velocities of 50-150 km/s. Given the very small spatial diameters of
type II spicules, it is quite possible that previous observations with
temporal and spatial resolution of order 30 s and 0.5-1\arcsec have
missed most of the dynamics of the type II spicules, and have instead
interpreted a superposition of many type II spicules as one spicule,
or perhaps even as a background chromosphere.

What is the impact of all of these spicules on the transition region
and corona? How do the different types of spicules impact the TR? Do
type I spicules just move the preexisting TR emission up and down
\citep[see, e.g.][]{deWijn+DePontieu2006}, or are they also associated
with heating to TR temperatures (perhaps less likely)?  Does a
significant fraction of type II spicules get heated to 100,000 K as is
suggested by our comparison with TRACE C IV data? Are type II spicules
the dominant source of heating in the magnetized chromosphere (or even
corona)? Are type II spicules signs of chromospheric evaporation?
Does a small but significant fraction of the type II spicules get
heated to coronal temperatures?  These are some of the unresolved
questions which are of great importance since they may well shed light
on the dominant heating mechanisms of both the chromosphere and
corona, as suggested previously by \citet{Ath82,Ath00,Wit83,Bud98}.
The access to extremely high resolution Hinode data has made the
resolution of these issues within view, for the first time.
Comparisons with high-resolution EUV and UV spectra and imaging will
help reveal whether reconnection dominates the jets visible in UV/EUV
\citep{Wil00} or whether, for example, electron beam heating or
conductive heating from the corona has any role to play in the
formation of type II spicules.  For example, such UV data has been
used recently by \citet{McIntosh+etal2006,McIntosh+etal2007} to
suggest that reconnection-driven jets can explain the spatial
distribution on supergranular scales of redshifts and blueshifts in UV
spectral lines that are formed at TR temperatures.

Our observations indicate that both wave-driven jets and
reconnection-driven jets are prevalent in the chromosphere. The
relative importance of these mechanisms seems to vary significantly
between active regions, quiet Sun and coronal holes.  Many of these
jets undergo vigorous transverse motions that are caused by
\alfven~waves.  Preliminary estimates suggest that these \alfven~waves
carry an energy flux that may be of importance for the local energy
balance, and once they reach the corona, can play a significant role
in the heating of the quiet Sun corona and acceleration of the solar
wind \citep{DePontieu+etal2007c}. The role of Alfv\'en waves in the
formation of spicules or the solar wind has been previously discussed
by, e.g., \citet{Hollweg+etal1982,Kud99}, although such waves had not
been previously observed in spicules.

%
{\em We are grateful to the Hinode team for their efforts in the
  design, building and operation of the mission. Hinode is a Japanese
  mission developed and launched by ISAS/JAXA, with NAOJ as domestic
  partner and NASA and STFC (UK) as international partners. It is
  operated by these agencies in co-operation with ESA and NSC
  (Norway). SOT was developed jointly by NAOJ, LMSAL, ISAS/JAXA, NASA,
  HAO and MELCO. B.D.P.  was supported by by NASA contracts
  NNG06GG79G, NNG04-GC08G, NAS5-38099 (TRACE) and NNM07AA01C (HINODE).
  SWM was supported by grants from the NSF (ATM-0541567) and NASA
  (NNG05GM75G, NNG06GC89G).}


\clearpage

\end{document}